\documentclass[3p,times,procedia]{elsarticle}
\usepackage{nupha_ecrc}

\volume{00}

\firstpage{1}

\journalname{Nuclear Physics A}

\runauth{}

\jid{nupha}

\jnltitlelogo{Nuclear Physics A}

\usepackage{amssymb}
\usepackage{amsmath, amssymb, graphics, setspace}

\usepackage[figuresright]{rotating}

\begin{document}

\begin{frontmatter}



\dochead{XXVIIth International Conference on Ultrarelativistic Nucleus-Nucleus Collisions\\ (Quark Matter 2018)}

\title{From in-Medium Color Forces to Transport Properties of QGP}


\author{Shuai Y.F. Liu and Ralf Rapp}

\address{Cyclotron Institute and Department of Physics and Astronomy, Texas A{\&}M University, 
	College Station, TX 77843-3366, USA}

\begin{abstract}
A thermodynamic quantum many-body $T$-matrix approach is employed to study the spectral and transport 
properties of the quark-gluon plasma at moderate temperatures where nonperturbative effects
are essential. For the partonic two-body interaction we utilize a QCD-inspired 
Hamiltonian whose color forces are motivated by the heavy-quark (HQ) limit including 
remnants of the confining force, and augmented by relativistic corrections. We solve the in-medium 
parton propagators and $T$-matrices selfconsistently and resum the skeleton diagrams 
for the equation of state (EoS) to all orders thereby accounting for the dynamical
formation of two-body bound states. Two types solutions for the in-medium potential are found in 
when fitting to lattice-QCD data for the EoS, HQ free energy and quarkonium correlators: 
a weakly-coupled scenario (WCS) with a (real part of the) potential close to the free energy, 
resulting in moderately broadened spectral functions and weak bound states near $T_c$, 
and a strongly-coupled scenario (SCS), with a much stronger potential which produces large 
imaginary parts (``melting" the parton spectral functions) and generates strong bound states 
near $T_c$. We calculate pertinent transport coefficients (specific shear viscosity and HQ 
diffusion coefficient) and argue that their constraints from heavy-ion phenomenology 
unambiguously favor the strongly-coupled scenario.  

\end{abstract}


\end{frontmatter}


\section{Introduction}
\label{sec_intro}
Anderson emphasized in his famous paper ``More is different"~\cite{Anderson:1972pca}, that
``the ability to reduce everything to simple fundamental laws does not imply the ability to 
start from those laws and reconstruct the Universe". This means that a quantum many-body theory for 
``reconstructing the Universe" is as critical an ingredient as the fundamental laws 
themselves. The quark-gluon plasma (QGP) created in heavy-ion collisions provides a unique
opportunity to study the quantum many-body theory of hot QCD. The complexity of this 
system renders exact/controlled first-principles calculations for several quantities 
of interest challenging. However, one might expect 
that the QGP, as a quantum many-body system, can be characterized by ``universal" properties 
that do not depend on precise details of the underlying interaction;
if so, simplified  QCD-inspired models should be a useful tool to gain insights into the 
structure of the QGP. 
The $T$-matrix approach employed in this work is an example for this philosophy.
While the partonic input Hamiltonian is adopted from a relatively simple Cornell-type potential, 
with medium effects quantitatively constrained by three sets of lattice-QCD data
(equation of state (EoS), heavy-quark (HQ) free energy and euclidean quarkonium correlators),
the many-body theory is rather elaborate, with selfconsistent one- and two-body Green's functions, 
and a resummed skeleton diagram series for the Luttinger-Ward-Baym 
functional~\cite{Liu:2016ysz,Liu:2017qah} retaining their full off-shell properties.
This allows to investigate how basic properties 
of the underlying force manifest themselves in a wide range of QGP properties~\cite{Liu:2016ysz,Liu:2017qah,Liu:2018syc}, 
in particular microscopically emerging spectral functions and transport coefficients 
which are not part of the fit and thus a prediction of the approach. 

In the following, we briefly outline the theoretical setup of the $T$-matrix approach 
(Sec.~\ref{sec_tm}), discuss the main results and physical insights (Sec.\ref{sec_res}), 
and conclude (Sec.~\ref{sec_con}). 

\section{Theoretical formalism for \textit{T}-matrix approach}
\label{sec_tm}
The  ``fundamental" degrees of freedom and their interactions are characterized by an in-medium 
effective Hamiltonian~\cite{Liu:2016ysz,Liu:2017qah}.
\begin{align}
&H=\sum\varepsilon(\textbf{p})\psi^\dagger(\textbf{p})\psi (\textbf{p})+
\frac{1}{2}\psi^\dagger(\frac{\textbf{P}}{2}-\textbf{p})\psi^\dagger(\frac{\textbf{P}}{2}+\textbf{p})
V \psi(\frac{\textbf{P}}{2}+\textbf{p}')\psi(\frac{\textbf{P}}{2}-\textbf{p}') \ , 
\label{eq_Hqgp}               
\end{align}
with parton energies $\varepsilon(\textbf{p})=\sqrt{M^2+\textbf{p}^{2} }$ whose masses are 
parameters that are constrained by the lQCD EoS. The two-body potential is composed of 
a color Coulomb ($ V_{\mathcal{C}} $) and confining ($ V_{\mathcal{S}} $) terms,  
\begin{align}
V(\textbf{p},\textbf{p}')=\mathcal{F}^\mathcal{C}R^\mathcal{C}V_\mathcal{C}(\textbf{p}-\textbf{p}')
+\mathcal{F}^\mathcal{S}R^\mathcal{S}V_\mathcal{S}(\textbf{p}-\textbf{p}') \ .
\label{eq_vp}
\end{align} 
withi color Casimir factors, $F^{\mathcal{C}(\mathcal{S})}$, and relativistic corrections, 
$ R^{\mathcal{C}(\mathcal{S})}$.
Its static limit in coordinate space, $ \tilde{V}(r) $, is directly related to the HQ free energy 
as computed in lQCD,
\begin{align}
&F_{Q\bar{Q}}(r,\beta)=\frac{-1}{\beta}\ln \bigg[\int_{-\infty}^{\infty} 
dE\,e^{-\beta E} \frac{-1}{\pi}\text{Im}[\frac{1}{E+i\epsilon-\tilde{V}(r)
	-\Sigma_{Q\bar Q}(E+i\epsilon,r)}]\bigg] \ , 
\label{eq_FreeEfinal}               
\end{align}
where $ \Sigma_{Q\bar Q}(E+i\epsilon,r) $ is the two-body selfenergy 
whose $ r $ dependence encodes interference effects~\cite{Liu:2017qah}.

With this Hamiltonian, a selfconsistent Brueckner $T$-matrix approach is carried out. 
The ladder resummation is given by
\begin{align}
&T(E,\textbf{p},\textbf{p}')=V(\textbf{p},\textbf{p}')+
\int_{-\infty}^{\infty}\frac{d^3\textbf{k}}{(2\pi)^3}V(\textbf{p},\textbf{k})
G^{0}_{(2)}(E,\textbf{k})T(E,\textbf{k},\textbf{p}') 
\label{eq_T}
\end{align}
with the in-medium 1- and 2-body propagators in Matsubara representation, 
$ G=([G^{0}]^{-1}-\Sigma)^{-1}$ and 
$G^{0}_{(2)}=-\beta^{-1}\sum_{\omega_n} G(iE_n-\omega_n)G(-i\omega_n)$, respectively. The 
single-parton selfenergy,
\begin{equation}
\Sigma=[G^0]^{-1}-G^{-1}=\int d\tilde{p}~\,T(G) G\equiv-\beta^{-1}\sum_{\nu_n}
\int \frac{d^{3}\textbf{p}}{(2\pi)^3}T(i\omega_{n}+i\nu_{n})G(i\nu_n) \ , 
\label{eq_selfE} 
\end{equation}
is a nonlinear functional of $ G $, characterizing the selfconsistency and satisfying 
the conservation laws~\cite{Baym:1961zz,Baym:1962sx}; it can also be derived by a functional 
derivative of Luttinger-Ward functional (LWF)~\cite{PhysRev.118.1417}, $\Phi$, 
as $ \delta\Phi/\delta G=\Sigma=\int d\tilde{p}TG $.  For the latter we resum the ladder diagrams 
to infinite order using a newly implemented matrix-log method to handle the extra $ 1/\nu $ factor, 
as
\begin{align}
\Phi=\frac{1}{2}\sum\text{Tr} &\bigg\{G\bigg[V+\frac{1}{2}V G^{0}_{(2)}V+\ldots
+\frac{1}{\nu}VG^{0}_{(2)}V G^{0}_{(2)}\ldots .V+\ldots\bigg]G\bigg\} 
= -\frac{1}{2}\ln[1-VG^{0}_{(2)}] \,.
\label{eq_phi2}
\end{align}
The grand-potential ($ \Omega=-P $) with selfconsistent propagators thus takes the 
form~\cite{PhysRev.118.1417,Baym:1962sx}
\begin{equation}
\Omega = \mp\sum\text{Tr}\{\ln(-G^{-1})+[(G^0)^{-1}-G^{-1}] G\}\pm\Phi \ ,
\label{eq_Omega}
\end{equation}
essentially corresponding to a single-particle and an interaction ($\Phi$) contribution. 
Our fit of this part of the formalism to the lQCD EoS essentially constrains the effective 
light-quark and gluon masses of the Hamiltonian. 
On the other hand, lQCD data for the HQ free energy largely constrain the in-medium
potential.

Since our approach is evaluated in real-time, spectral functions and scattering amplitudes can 
be extracted from the parton propagators and $T$-matrices, yielding direct insights into QGP 
structure. They can be further used to compute transport coefficients within the same framework. 
The shear viscosity is calculated by a Kubo formula at dressed one-loop level~\cite{Liu:2016ysz},
\begin{equation}
\eta=-\sum_i\pi d_i \int
\frac{d^3\textbf{p}d\omega}{(2\pi)^3} \frac{p_x^2p_y^2}{\varepsilon^2_i(p)}
\rho^2_i(\omega,p)\frac{d n_i(\omega)}{d\omega} \ ,
\end{equation}
where $ d_i $, $ n_i (\omega) $, and $ \rho_{i} $ are the parton degeneracies, thermal
distribution and spectral functions, respectively.
The charm-quark friction coefficient and corresponding spatial diffusion
coefficient are obtained by an off-shell generalization~\cite{Liu:2016ysz,Liu:2018syc} 
of previous $ T $-matrix calculations~\cite{Riek:2010fk,Prino:2016cni}, 
\begin{align}
A_c(p)=&\left\langle (1-\frac{\textbf{p}\cdot\textbf{p}'}{p^2})\rho_i\rho_i\rho_c\right\rangle \ ,
\quad D_s  = T/(A_c(0)M_c)  \ .  
\end{align}

\section{Numerical Results and Insights}
\label{sec_res}

\begin{figure}[b]
	\centering
	\includegraphics[width=0.245\columnwidth]{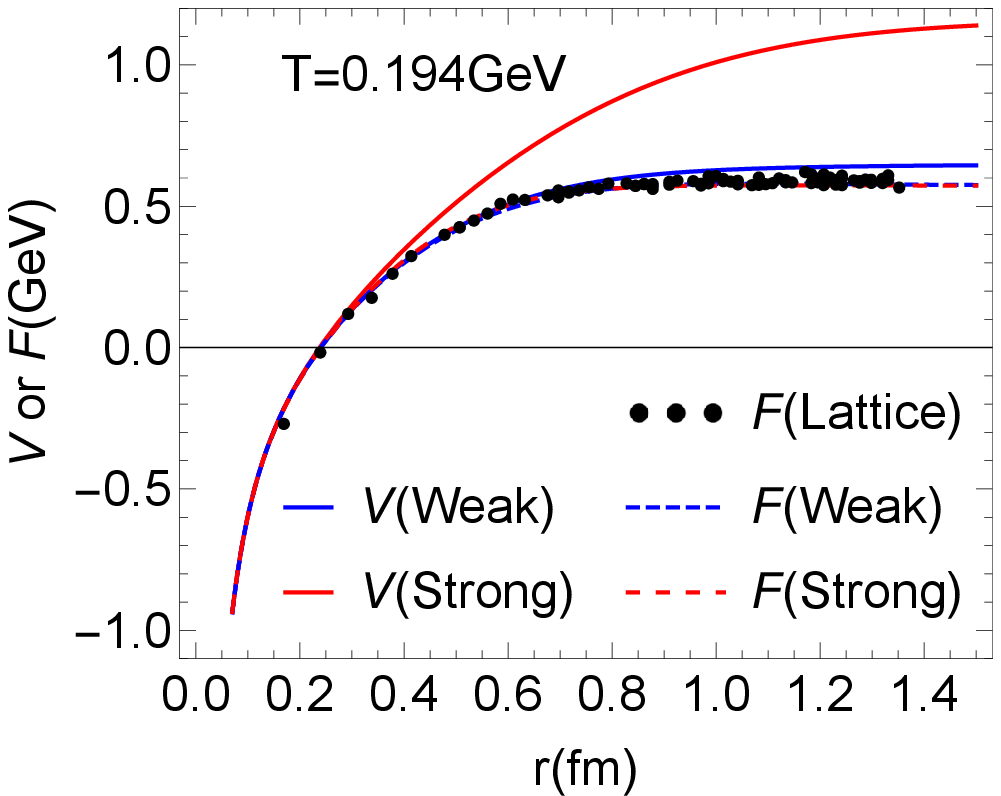}
	\includegraphics[width=0.245\columnwidth]{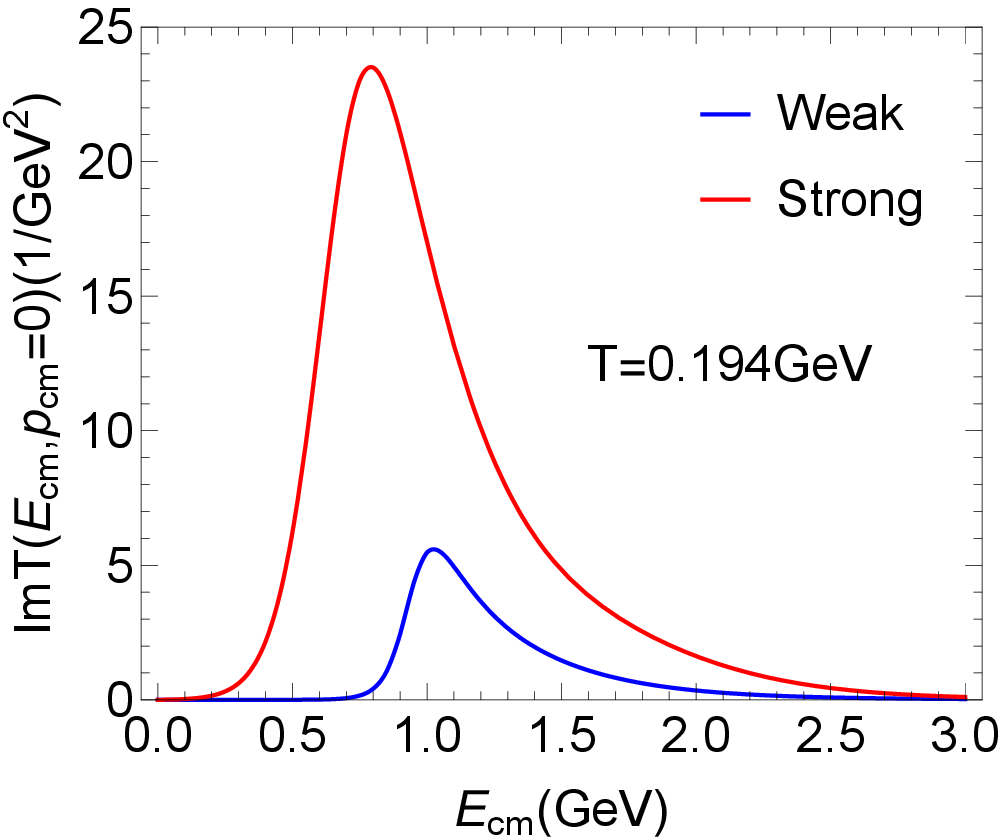}
	\includegraphics[width=0.233\columnwidth]{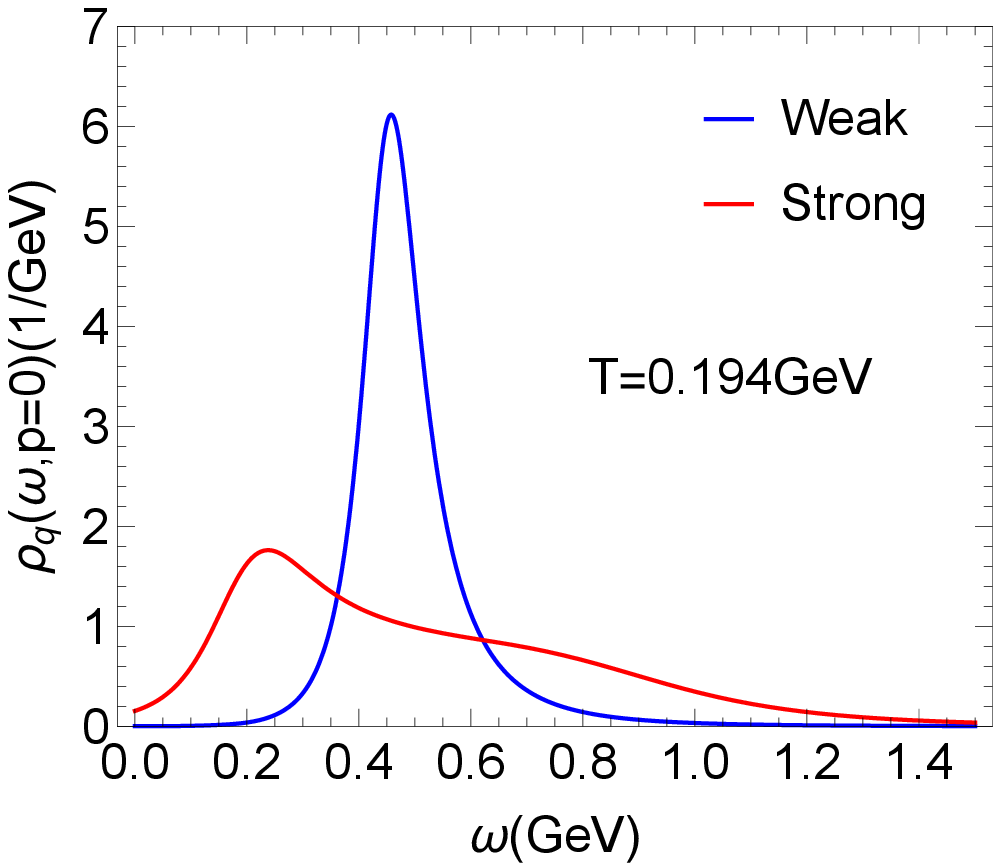}
	\includegraphics[width=0.253\columnwidth]{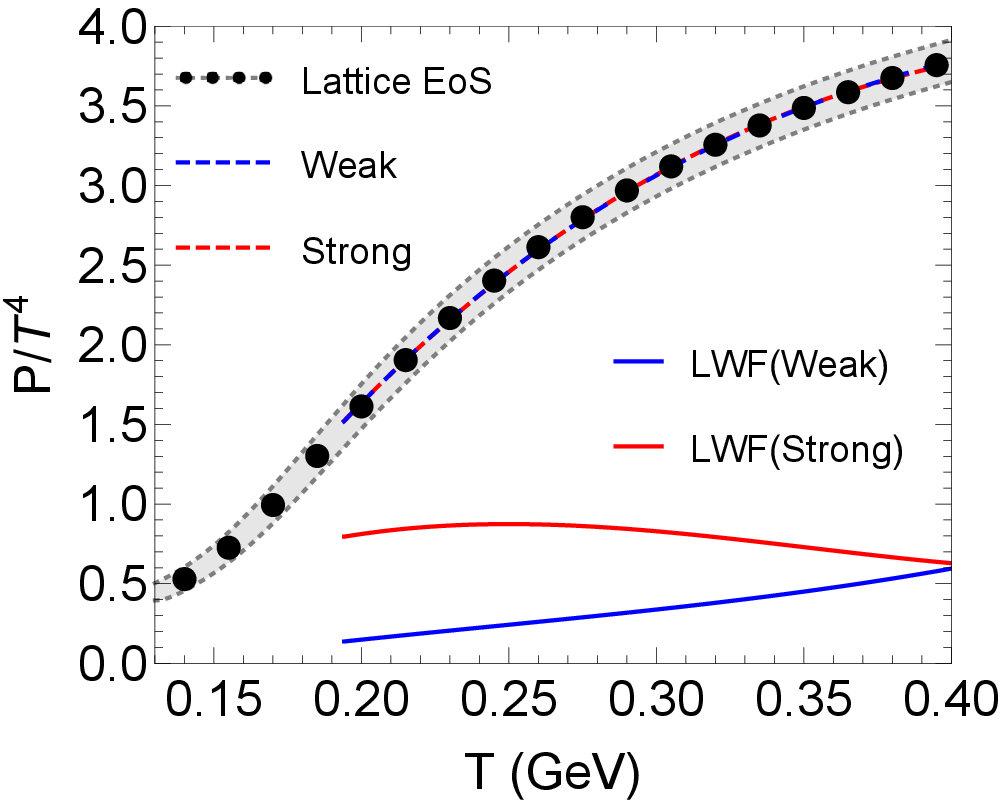}
	\vspace{-0.3cm}
	\caption{From left to right: potential and HQ free energy, imaginary part of color-singlet
$T$-matrix, light-quark spectral function and pressure with LWF contribution, for the SCS (red lines) 
and WCS (blue lines).}
	\label{fig_micro}
\end{figure}
\begin{figure} [t]
	\centering
	\includegraphics[width=0.35\columnwidth]{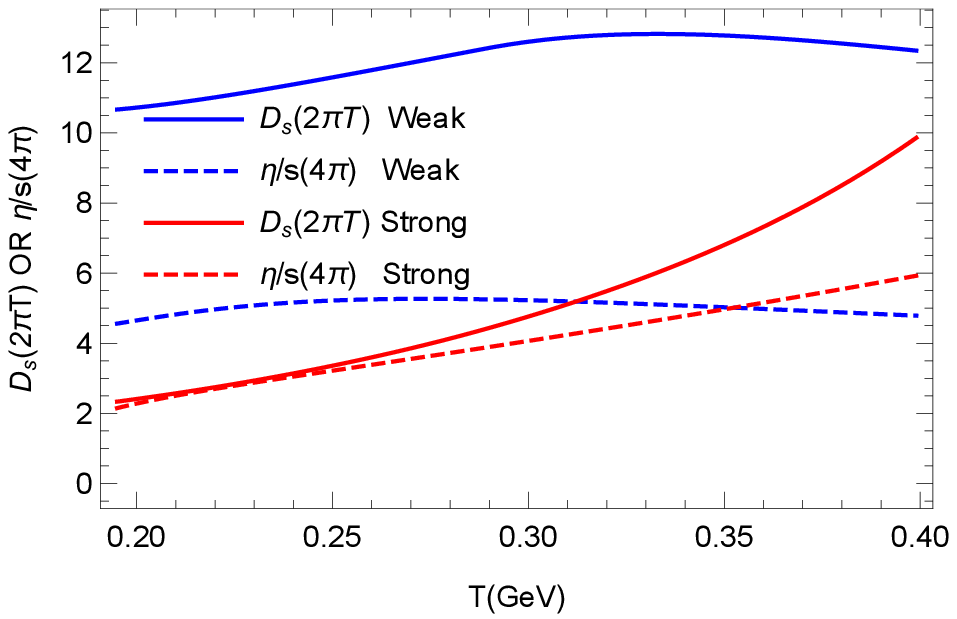}\hspace{0.5cm}
	\includegraphics[width=0.35\columnwidth]{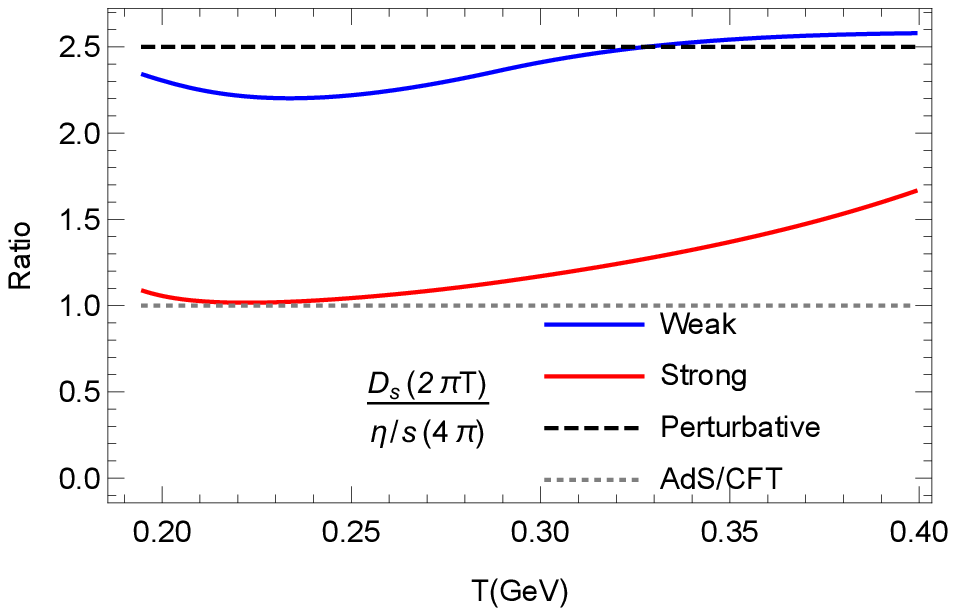}
	\vspace{-0.3cm}
	\caption{Charm-quark diffusion coefficient and specific shear viscosity (left)
and their ratio (right) in the SCS (red lines) and WCS (blue lines)~\cite{Liu:2016ysz,Liu:2017lhv}.}
	\label{fig_trans}
\end{figure}

It turns out that the selfconsistent fits to the EoS and quarkonium correlators and free energies
support two types of solutions: a strongly-coupled (SCS) and weakly-coupled (WCS) 
scenario~\cite{Liu:2017qah,Liu:2018syc}. However, the underlying micro-physics for the two 
solutions is very different, as illustrated in Fig.~\ref{fig_micro}.
The SCS features a long-range remnant of the confining potential that is much larger than the free 
energy (left panel); this leads to strong resonances in the $ T $-matrix (second-from-left panel) 
which in turn induce large parton collision rates which melt the quasiparticle peaks in their 
spectral functions (second from right panel) signaling a transition in the degrees of freedom. 
This transition can also be seen in the pressure, where the LWF ($\Phi$), encoding the
bound-state contribution accounts for more than 50\% close to $T_c$. 
On the other hand, in the WCS the potential is close to the free energy; this leads to relatively 
weak resonance correlations near $T_c$ and small parton collision rates so that their spectral 
functions retain well-defined quasiparticle peaks. For the pressure, the  LWF contribution remains 
small at all temperatures, with no indication for a transition in the degrees of freedom.

How can we distinguish these solutions? It turns out that the transport parameters of the two 
scenarios are quite different, cf.~Fig.~\ref{fig_trans}. For the SCS, the specific shear 
viscosity is about twice the conjectured lower bound~\cite{Kovtun:2004de}, but another factor 
of $\sim$2 larger in the WCS at low $T$. The difference in the HQ diffusion coefficient is more 
pronounced: it is about twice the thermal wavelength in the SCS, but up to another factor of 5 
larger in the WCS. From a phenomenological point~\cite{Prino:2016cni,Rapp:2018qla,Liu:2018syc}, 
this clearly favors the SCS. Of particular interest is the ratio of $D_s(2\pi T)$ over
$4\pi\eta/s$~\cite{Rapp:2009my}, which is expected to be near 1 in the strongly coupled 
limit~\cite{Gubser:2006bz}, but 5/2 in a perturbative system~\cite{Danielewicz:1984ww}. 
The latter is indeed realized in our WCS at all temperatures, while the former is realized 
in the SCS at low temperatures, and increasing toward higher temperatures.
This corroborates the classification of the two scenarios as ``strongly" and ``weakly" coupled.  


\section{Conclusion and Discussion}
\label{sec_con}
We have utilized a thermodynamic $T$-matrix approach to understand and connect various properties 
of the QGP in the nonperturbative regime. We have employed a QCD-inspired effective Hamiltonian and
constrained its parameters using lattice-QCD data on the equation of state, heavy-quarkonium 
correlators and free energies. Carrying out a full quantum calculation in fitting these quantities, 
the resulting spectral and transport properties of two solution types -- SCS and WCS -- exhibited
a ``sufficient and necessary" correlation link: ``large color (string) potential" $\Leftrightarrow $
``strong two-body resonances" $\Leftrightarrow $ ``broad (non-quasiparticle) spectral functions" 
$ \Leftrightarrow $ ``small viscosity/spatial diffusion coefficient". Considering the phenomenological 
constraints for the transport coefficients from hydrodynamic~\cite{Bernhard:2016tnd,Niemi:2018ijm} 
and open heavy-flavor data in heavy-ion collisions~\cite{Prino:2016cni,Rapp:2018qla,Liu:2018syc}, 
this chain implies that the microscopic structure of the QGP should be close to the one predicted 
by the SCS. The basic feature of the underlying force is a long-range remnant of the confining 
potential, which generates a strong-coupling behavior for the long-wavelength excitations of the 
system, {\it i.e.}, its low-momentum spectral functions and transport properties, while recovering 
weakly-coupled behavior at short distance. Ultimately, these are manifestation of the nontrivial 
vacuum structure of QCD and its running coupling that persist into the QGP. 
Our conclusions here are different from that of Bayesian analysis on lQCD Wilson line/loop 
data~\cite{Burnier:2014ssa} whose results for the extracted potential and imaginary parts
are close to the WCS in the $T$-matrix approach; further efforts are required to disentangle
this apparent discrepancy.

\section*{Acknowledgments}
This work was supported by the U.S.~National Science Foundation (NSF) under 
grant no.~PHY-1614484.



\bibliographystyle{apsrev4-1}
\bibliography{refcnew}







\end{document}